\documentclass[a4paper,12pt]{article}
\pdfoutput=1 % if your are submitting a pdflatex (i.e. if you have
\usepackage{jheppub} % for details on the use of the package, please
\usepackage[T1]{fontenc} % if needed
\usepackage{lmodern}

\usepackage{tikz}

\let\oldFootnote\footnote
\newcommand\nextToken\relax

\renewcommand\footnote[1]{%
    \oldFootnote{#1}\futurelet\nextToken\isFootnote}

\newcommand\isFootnote{%
    \ifx\footnote\nextToken\textsuperscript{,}\fi}

\usepackage{enumerate}

\def\id{{1 \kern-.28em {\rm l}}}

\def\K3{{\bf K3}}
\def\journal#1&#2(#3){\unskip, \sl #1\ \bf #2 \rm(19#3) }
\def\andjournal#1&#2(#3){\sl #1~\bf #2 \rm (19#3) }

\def\bar{\overline}

\def\ie{{\it i.e.}}
\def\eg{{\it e.g.}}

\def\etc{{\it etc}}

\def\frac#1#2{{#1\over#2}}

\def\inbar{\,\vrule height1.5ex width.4pt depth0pt}
\def\IC{\relax\hbox{$\inbar\kern-.3em{\rm C}$}}
\def\IR{\relax{\rm I\kern-.18em R}}
\def\IP{\relax{\rm I\kern-.18em P}}

\catcode`\@=11
\def\slash#1{\mathord{\mathpalette\c@ncel{#1}}}
\def\MM{{\cal M}}
\def\NN{{\cal N}}

\def\underrel#1\over#2{\mathrel{\mathop{\kern\z@#1}\limits_{#2}}}

\catcode`\@=12

\def\exp{{\rm exp}}

\def\ie{{\it i.e.}}
\def\eg{{\it e.g.}}

\title{Wilson loop in a $T\bar{T}$ like deformed $\rm{CFT}_2$}

\author{Soumangsu Chakraborty}
\affiliation{Racah Institute of Physics,\\The Hebrew University,\\ Jerusalem 91904, Israel}
\abstract{In this paper we study string theory in the background $\mathcal{M}_3$ that interpolates between $AdS_3$ in the IR and linear dilaton spacetime $\mathbb{R}^{1,1}\times\mathbb{R}_\phi$ in the UV. Via holographic duality this background corresponds to $\rm{CFT}_2$ deformed by a dimension $(2,2)$ operator. Here we discuss the holographic Wilson loop in such a model and shed more light in support of the non-local structure of the theory (Little String Theory (LST)) in the UV. We also discuss quantum and thermal phase transitions of the boundary theory.}

\begin{document}
\maketitle
\flushbottom

\section{Introduction}

Recently there has been a considerable interest \cite{Giveon:2017nie,Giveon:2017myj,Asrat:2017tzd,Chakraborty:2018kpr} in the study of string theory in the background
\begin{eqnarray}\label{theback}
\MM_3\times \NN,
\end{eqnarray}
where the background $\MM_3$ interpolates between linear dilaton geometry $\mathbb{R}^{1,1}\times\mathbb{R}_\phi$ in the ultraviolet (UV) and $AdS_3$ in the infrared (IR) and $\NN$ is a 7-dimensional compact space.

From UV perspective this background can be realized as the bulk description of certain 2-dimensional vacua of LST with $N\gg 1$ fundamental strings \cite{Aharony:1998ub,Giveon:1999zm}. While the UV physics is governed by that of the underlying LST, the theory approaches a $\rm{CFT}_2$  dual to $AdS_3$ in the bulk, in the IR. The $AdS_3$ geometry in the IR corresponds to the near-horizon geometry of the fundamental strings in the linear dilaton background.

However, from the IR perspective, these backgrounds can be realized as an irrelevant deformation of the $\rm{CFT}_2$ dual to the above $AdS_3$ by a certain class of dimenision $(2,2)$ quasi-primary \cite{Giveon:2017nie,Giveon:2017myj,Asrat:2017tzd,Chakraborty:2018kpr}. Such a deformation of the IR $\rm{CFT}_2$ involves flowing up the renormalization group (RG) and hence may appear to be ambiguous and ill-defined. However, as has been argued in \cite{Giveon:2017nie,Giveon:2017myj,Asrat:2017tzd,Chakraborty:2018kpr} (both from worldsheet and spacetime point of view), the situation here is much better and one can actually flow up the RG. 

In this paper we are going to investigate the behaviour of the holographic Wilson loop operator of the dual boundary theory all along the RG. Since the theory in the UV is non-local, we expect the Wilson loop operator to exhibit non field-theoretic behaviour at short distances. We would investigate the non-locality scale and comment on quantum and thermal phase transitions of the dual spacetime theory. 

As we will discuss, in details, in the next section, the large $T_E$ behavior of the expectation value of the Wilson loop operator, $W$, is given by (see \eg\ \cite{Maldacena:1998im})
\begin{eqnarray}\label{eWL}
\langle W\rangle\sim e^{T_EE(L)},
\end{eqnarray}
where $T_E$ and $L$ are the lengths of the sides of a rectangular loop (residing on the boundary) with edges parallel to the direction of Euclidean time and space respectively and $E$ is the potential energy of the quark anti-quark pair separated by a distance $L$. For theories with two fixed points, $E(L)$ interpolates between quark anti-quark potential in the UV CFT to that in the IR CFT. From the behaviour of $E$ as a function of $L$ one can understand the nature of transition from the UV to the IR and determine the different phases of the theory. A natural question to ask is what happens to $E(L)$ if the short distance physics is non-local and not governed by a UV fixed point \cite{Giveon:2017nie,Giveon:2017myj,Asrat:2017tzd,Chakraborty:2018kpr}.

As discussed in \cite{Giveon:2017nie,Giveon:2017myj,Asrat:2017tzd,Chakraborty:2018kpr}, string theory in $\MM_3\times \NN$ is closely related to $T\bar T$ deformed $\rm{CFT}_2$. Following the recent works of \cite{Smirnov:2016lqw,Cavaglia:2016oda}, there has been a considerable progress in understanding different aspects of $T\bar T$ deformed $\rm{CFT}_2$; see \eg\ \cite{McGough:2016lol,Shyam:2017znq,Giribet:2017imm,Kraus:2018xrn,Cardy:2018sdv,Cottrell:2018skz,Aharony:2018vux,Dubovsky:2018dlk,Bonelli:2018kik,Taylor:2018xcy,Datta:2018thy,Donnelly:2018bef,Babaro:2018cmq,Conti:2018jho,Chen:2018eqk,Hartman:2018tkw,Aharony:2018bad} and \cite{Guica:2017lia,Bzowski:2018pcy,Chakraborty:2018vja,Apolo:2018qpq,Aharony:2018ics} for related works. It is natural to ask if the expectation value of the Wilson loop operator in these models is similar to that of the string background $\MM_3\times \NN$. But unfortunately, there are no simple method to calculate the expectation value of Wilson loop operator in strongly coupled gauge theories. We will therefore assume the duality between the string theory in $\MM_3\times \NN$ and  $T\bar T$ deformed $\rm{CFT}_2$ and investigate the properties of the Wilson loop operator of  $T\bar T$ deformed $\rm{CFT}_2$  via holography.

The plan of this note is as follows. In section \ref{sec2} we give a brief review of the necessary aspects of the construction in \cite{Giveon:2017nie,Giveon:2017myj,Asrat:2017tzd,Chakraborty:2018kpr} and previous works on holographic Wilson loop operator \cite{Maldacena:1998im}. In section \ref{sec3} we compute the expectation value of the holographic Wilson loop operator and the potential energy of the quark anti-quark system  separated by a distance $L$ at zero temperature. We compute its large and small $L$ behaviour and estimate the non-locality scale of the theory in the UV. We find that the theory exhibits second order quantum phase transition.

In section \ref{sec4} we generalize the holographic results  to finite temperature and discuss thermal and quantum phase transitions in the theory. In section \ref{sec5} we discuss our results and compare ours with those obtained in \cite{Chakraborty:2018kpr} and comment on  few possible avenues for future works.

\section{Review} \label{sec2}

\subsection{An irrelevant deformation of $AdS_3/CFT_2$ }

The authors of \cite{Smirnov:2016lqw,Cavaglia:2016oda} have shown that a certain irrelevant deformation of a generic $\rm{CFT}_2$ by an operator which is bilinear in the stress tensor (to be precise $T\bar T$) is, to a large extent, solvable. For example one can calculate the exact spectrum of the theory. At high energy, the theory appears to be well defined in spite of the fact that this involves flowing up the RG. The entropy of the system interpolates between that of a $\rm{CFT}_2$ in the IR (Cardy entropy) and one with Hagedorn entropy at very high energies. Thus the short distance behaviour of the theory is not governed by local QFT. In the case of supersymmetric theories, the deforming operator being the top component of the superfield, preserves supersymmetry. Inspired by the results of \cite{Smirnov:2016lqw,Cavaglia:2016oda}, the authors of \cite{Giveon:2017nie,Giveon:2017myj,Asrat:2017tzd,Chakraborty:2018kpr} discussed a string model that shares many properties with the $T\bar T$ deformed $\rm{CFT}_2$.

In those particular cases where the IR $\rm{CFT}_2$ has a holographic dual in $AdS_3$, the $T\bar T$ deformed $\rm{CFT}_2$ studied in \cite{Smirnov:2016lqw,Cavaglia:2016oda} corresponds to a double trace deformation of the original duality. Generic double trace deformations of $AdS/CFT$ dualities are studies in \cite{Witten:2001ua,Berkooz:2002ug}. The authors of \cite{Giveon:2017nie} made an observation that there exists an irrelevant single trace deformation
\begin{eqnarray}\label{D(x)}
\delta\mathcal{L}=\mu D(x),
\end{eqnarray}
 of the boundary $\rm{CFT}_2$ that does the same job as the $T\bar T$ deformation of $\rm{CFT}_2$ where $\mu$ is the irrelevant coupling of spacetime theory with mass dimension $(-1,-1)$ and $D(x)$ is a certain dimension $(2,2)$ quasi-primary of the spacetime Virasoro. The construction of the operator $D(x)$ is discussed in \cite{Kutasov:1999xu}.  Despite being irrelevant, this single trace deformation $D(x)$ of the spacetime $\rm{CFT}_2$ is under control as it induces on the worldsheet a truly marginal deformation of the form \footnote{See \cite{Israel:2003ry} for similar deformations of WZW models on $AdS_3$.}
\begin{eqnarray}\label{JJbar}
\delta\mathcal{L}_{ws}=\lambda J^-\bar{J}^-,
\end{eqnarray}
where $J^-$ and $\bar{J}^-$ are the holomorphic and anti-holomorphic components of the null $SL(2,\mathbb{R})$ currents on the worldsheet whose zero modes gives rise to the boundary Virasoro generators $L_{-1}$ and $\bar{L}_{-1}$ respectively \cite{Giveon:1998ns}. The coupling $\lambda$ on the worldsheet is truly marginal as it has worldsheet dimension $(0,0)$ and is related to the spacetime coupling $\mu$ by some dimensionfull constant. The deformation $D(x)$ is in some sense universal because any $\rm{CFT}_2$ with $AdS_3$ dual and NS-NS B-field flux with support in $AdS_3$ contains such a deformation.

The current-current deformation (non Abelian Thirring) \eqref{JJbar} of the worldsheet sigma model is exactly solvable. Using standard worldsheet techniques, one can read off the  deformed metric, dilaton and B-field which we refer to as $\MM_3$. This deformed background is given by \cite{Forste:1994wp}
\begin{eqnarray}
ds^2&=&f^{-1}\left(-dt^2+dx^2\right)+k\alpha'\frac{dU^2}{U^2},\nonumber \\
e^{2\Phi}&=&\frac{g_s^2}{kU^2}f^{-1},\label{background}\\
dB&=&\frac{2i}{U^2}f^{-1}\epsilon_3,\nonumber
\end{eqnarray}
where $f=1+\frac{1}{kU^2}$,  $k$ is the level of the worldsheet $SL(2,\mathbb{R})$ current algebra of the model and $g_s$ is the string coupling determined by the asymptotic value (\ie\ $U\to \infty$) of the dilaton field. 

The background $\MM_3$ in \eqref{background} interpolates between $AdS_3$ in the IR (\ie\ $U\to 0$) to linear dilaton geometry 
\begin{eqnarray}\label{LDg}
 \mathbb{R}^{1,1}\times \mathbb{R}_\phi
\end{eqnarray}
in the UV (\ie\ $U\to \infty$) where $\phi\sim \ln U$. Transition takes place at scales that depends of the coupling $\lambda$. Without loss of generality one can set the coupling to a convenient value as discussed in \cite{Giveon:2017nie,Giveon:2017myj}. 

As an example, the interpolating geometry $\MM_3$ is realized as follows. We start with a stack of $k$ NS5-branes  wrapped around a  four dimensional compact manifold (\eg\ $T^4$ or $K_3$). The near horizon geometry of the NS5-branes gives \eqref{LDg}. The string coupling $g_s$ goes to 0 near the boundary (\ie\ $U\to \infty$) where the dual field theory lives, but deep in the bulk (\ie\ $U\to 0$), $g_s\to\infty$. Now if we put $N$ fundamental strings (F1) stretched along $\mathbb{R}^{1,1}$, the resulting background is given by \eqref{background}. Upon addition of F1 strings, the IR geometry gets modified and the string coupling stops growing and saturates as $g_s^2\sim 1/N$. The smooth interpolation from linear dilaton background in the UV to $AdS_3$ in the IR corresponds to going from the near horizon geometry of the NS5 system to that of NS5+F1 system. The spacetime $\rm{CFT}_2$ in the IR has central charge $c=6kN$. The linear dilaton geometry in the UV describes a two dimensional vacua  of LST \cite{Aharony:1998ub} with Hagedorn density of states \cite{Aharony:2004xn} and diverging entropic c-function \cite{Chakraborty:2018kpr}. The inverse Hagedorn temperature is given by
\begin{eqnarray}
\beta_H=2\pi\sqrt{k\alpha'}.
\end{eqnarray}
If the original LST to start with is supersymmetric, the deformation preserves supersymmetry since F1 strings are BPS.

As discussed earlier, the single trace deformation \eqref{D(x)} of the spacetime $\rm{CFT}_2$ is not exactly same as the $T\bar{T}$ deformation as studies in \cite{Smirnov:2016lqw,Cavaglia:2016oda} but is closely related. To realize the difference let us consider a $\rm{CFT}_2$ that has a symmetric product form $\MM^N/S_N$ where $\MM$ is the $\rm{CFT}_2$ that forms the building block of the symmetric product orbifold CFT. In such a theory there are two natural choices of $T\bar{T}$ deformation namely $\sum_{i=1}^NT_i\sum_{j=1}^N\bar{T}_j$ and $\sum_{i=1}^NT_i\bar{T}_i$ where $T_i$ is the holomorphic component of the stress tensor of the $i^{th}$ block $\MM$ in the symmetric product. The first is the $T\bar{T}$ deformation of $\MM^N/S_N$ and is double trace, while the second is the  $T\bar{T}$ deformation of the block $\MM$ and is single trace. The spacetime $\rm{CFT}_2$ corresponding to string theory on the bulk $AdS_3$ is not exactly a symmetric product orbifold CFT but is closely related to it \cite{Argurio:2000tb,Giveon:2005mi}. The dual background corresponding to a certain Ramond vacuum (which preserves supersymmetry on a cylinder) of the boundary $\rm{CFT}_2$, is massless BTZ black hole. The strings and five branes that form the background are mutually BPS implying that their potential is flat. This in turn implies that there are continuum of states  corresponding to strings moving radially away from the five branes. Such states form a symmetric product CFT, as observed in matrix string theory \cite{Motl:1997th,Dijkgraaf:1997vv}. 

\subsection{Wilson loop}

In this subsection we give a brief review of the holographic Wilson loop operator; see \eg\  \cite{Maldacena:1998im} for details. The Wilson loop operators in any gauge theory are highly non-local operators in the theory and are defined by
\begin{eqnarray}\label{WL}
W(\mathcal{C})=\mathrm{Tr}\left[P\exp\left(i\oint_{\mathcal{C}}\mathcal{A}\right)\right],
\end{eqnarray}
where $\mathcal{C}$ is a closed loop  in spacetime where the gauge theory lives, $\mathcal{A}$ is the gauge field and $P$ denotes path ordering of the gauge connection along the contour $\mathcal{C}$. The physical meaning that one can associate with the Wilson loop is that of the phase factor that shows up in transporting a given external charged  particle in some representation along some closed path $\mathcal{C}$. The trace in \eqref{WL}, in principle, can be taken over any representation, but we will restrict ourselves only to the fundamental representation. The potential energy between a quark and an anti-quark pair in the gauge theory can be calculated from the expectation value of the Wilson loop operator $\langle W(\mathcal{C})\rangle$. To calculate $\langle W(\mathcal{C})\rangle$, let us first Wick rotate to Euclidean signature and then consider a rectangular loop with sides of length $T_E$ and $L$. Treating $T_E$ as the Euclidean time, one would expect large $T_E$ ($T_E\gg L$) behaviour of $\langle W(\mathcal{C})\rangle$ to be of the form $\langle W(\mathcal{C})\rangle\sim e^{-T_EE(L)}$ where $E(L)$ can be interpreted as the quark anti-quark potential energy. For large $N$ gauge theories at large 'tHooft coupling,  the complicated problem of computation of $\langle W\rangle$ is mapped to a classical problem of finding the minimal surface $\Sigma$ in the holographic bulk dual such that $\partial \Sigma=\mathcal{C}$.

The partition function in any quark system in any gauge theory is $\langle W\rangle$. At large $N$ and large 'tHooft limit, one can equate this to the classical string partition function $Z_{\rm{string}}[\partial \Sigma=\mathcal{C}]=e^{-S}$ associated with a single F1 string where $\Sigma$ is a genus zero compact worldsheet Riemann surface that minimizes the string action $S$ (Nambu-Goto or Polyakov) and satisfies $\partial \Sigma=\mathcal{C}$. In superstring theory, under S-duality a F1 string  gets mapped to a D1-brane, in that case one needs to replace the Nambu-Goto or Polyakov action in the string partition function by the Dirac-Born-Infeld (DBI) action .

\section{Bulk calculation: zero temperature}\label{sec3}

\subsection{Holographic Wilson loop}

\begin{figure}[h]
    \centering
    \includegraphics[width=.53\textwidth]{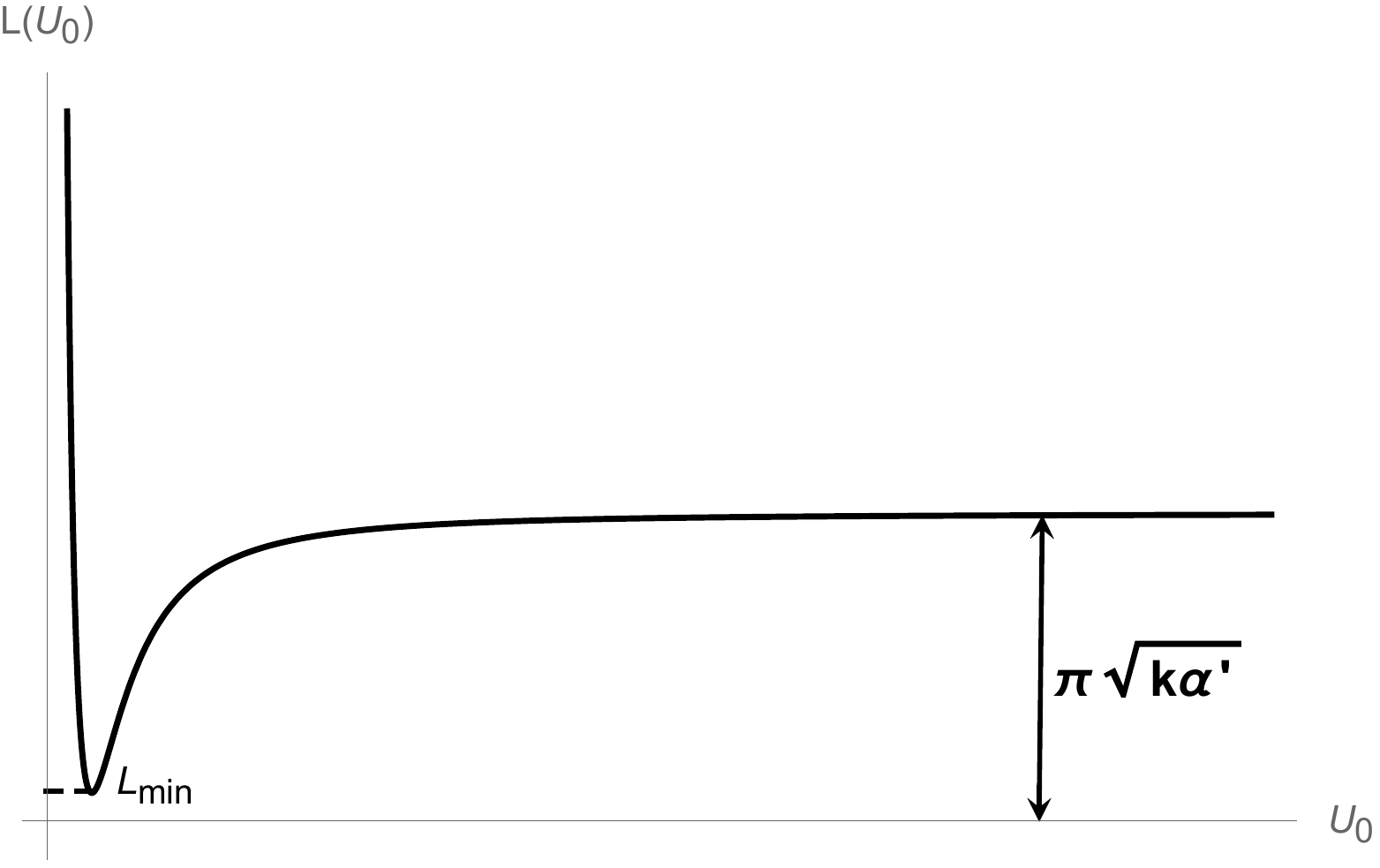}
    \caption{The quark anti-quark separation $L$ as a function of $U_0$ in $\mathcal{M}_3$ at zero temperature.}
    \label{M3 T=0 L vs U0}
\end{figure}

Let us consider a quark anti-quark pair at $x=L/2$ and $x=-L/2$ on the boundary of the background  manifold $\mathcal{M}_3$. Here "quark" means an infinitely massive W-boson connecting the stack of $k$ NS5-branes with the one 5-brane (on the boundary) which is far away from the stack.  The DBI action of the D1-brane hanging in the bulk with two of its ends fixed at the boundary, is given by
\begin{eqnarray}
S&=&\frac{1}{2\pi\alpha'}\int d\tau d\sigma e^{-\Phi}\sqrt{\text{det}\left(G_{MN}\partial_\alpha X^M\partial_\beta X^N\right)}\nonumber \\
&=&\frac{T_Ek}{2\pi\alpha'g_s}\int  dx\sqrt{\frac{U^4+\alpha'(kU^2+1)(\partial_x U)^2}{(kU^2+1)}}, \label{M3 T=0 action}
\end{eqnarray}
where $G_{MN}$ is the 10 dimensional target space metric and $X^M$s are the embeddings of the D1-brane in the target space and $(\tau,\sigma)$ are Euclidean coordinates on the worldvolume of the D1-brane. Here we choose to work in the static gauge, namely $\tau=t$  and $\sigma=x$. The equation of motion is given by
\begin{eqnarray}
\frac{U^4}{\sqrt{1+kU^2}\sqrt{U^4+\alpha'(1+kU^2)(\partial_x U)^2}}=\frac{U_0^2}{\sqrt{1+kU_0^2}},\label{M3 T=0 constrain}
\end{eqnarray}
where we have used the boundary condition that $U(x=0)=U_0$ and $\partial _x U|_{x=0}=0$.
This allows us to express the length of the separation of the two ends of the D1-brane, $L$, on the boundary as 
\begin{eqnarray}
L&=&2\sqrt{\alpha'}\int_1^{\infty}dy\frac{\sqrt{\left(\frac{1}{U_0^2}+ky^2\right)}}{y^2\sqrt{\frac{y^4\left(\frac{1}{U_0^2}+k\right)}{\left(\frac{1}{U_0^2}+ky^2\right)}-1}}\nonumber \\
&=&\frac{2\sqrt{\alpha'}}{U_0(1+kU_0^2)}\Bigg\{(1+kU_0^2)^{3/2}\mathcal{E}\left(\frac{-1}{1+kU_0^2}\right)-(2+kU_0^2)\sqrt{1+kU_0^2}\mathcal{K}\left(\frac{-1}{1+kU_0^2}\right)\nonumber \\
&&+(1+kU_0^2)^2\left(i\mathcal{K}(-1-kU_0^2)+\mathcal{K}(2+kU_0^2)\right)\Bigg\}, \label{M3 T=0 length}
\end{eqnarray}
where $y=\frac{U}{U_0}$, and $\mathcal{K}(m)$ and $\mathcal{E}(m)$ are complete elliptic integrals of the first kind and of the second kind respectively, given by
\begin{eqnarray}
\mathcal{K}(m)&=&\int_0^{\pi/2} d\theta \frac{1}{\sqrt{1-m\sin^2\theta}}=\frac{\pi}{2}~_2F_1\left[\frac{1}{2},\frac{1}{2};1;m^2\right],\nonumber \\
\mathcal{E}(m)&=&\int_0^{\pi/2} d\theta \sqrt{1-m\sin^2\theta}=\frac{\pi}{2}~_2F_1\left[-\frac{1}{2},\frac{1}{2};1;m^2\right].
\end{eqnarray}
In the deep $AdS_3$ regime (\ie\ $U_0\to 0$), $L$ behaves as
\begin{eqnarray}
L&= & \frac{2\sqrt{\alpha'}}{U_0}\left(\mathcal{E}(-1)+\mathcal{K}(2)-\frac{(2-i) \sqrt{\pi } \Gamma \left(\frac{5}{4}\right)}{\Gamma \left(\frac{3}{4}\right)}\right)+O(U_0)\approx \frac{1.19814}{U_0}\sqrt{\alpha'}+O(U_0).\nonumber \\
\end{eqnarray}
The $1/U_0$ dependence of $L$ is also observed in $AdS_5\times S^5$ as discussed in \cite{Maldacena:1998im}. In the linear dilaton regime (\ie\ $U_0\to \infty$), $L$ reaches a constant proportional to the string length $\sqrt{\alpha'}$.\footnote{This was first observed in one of the examples studied in \cite{Brandhuber:1998er}.} Large $U_0$ expansion of $L$ is given by
\begin{eqnarray}
L=\frac{\beta_H}{2}-\frac{\pi\sqrt{\alpha'}}{4\sqrt{k}U_0^2}+O\left(\frac{1}{U_0^4}\right). 
\end{eqnarray}
Figure (\ref{M3 T=0 L vs U0}) shows the plot of $L$ as a function of $U_0$.
Following are few comments on the $L$ vs $U_0$ plot:
\begin{enumerate}[{(i)}]
\item {As $L\to \infty$, $U_0\to 0$ implying that the bottom of the minimal surface goes deep inside the $AdS_3$ regime. In this regime the interaction between the quark anti-quark pair is governed by physics of the $AdS_3$ region of the full bulk geometry as depicted by RG.}

\item{In the linear dilaton regime (\ie\ $U_0\to \infty$), $L$ approaches the constant $\beta_H/2$. Here the spacetime physics is that of LST in UV. The separation of the two ends of the D1-brane is precisely same as that observed in the case of hairpin (or paperclip) branes in linear dilaton geometry (see \eg\ \cite{Lukyanov:2003nj,Brandhuber:1998er}). Fixed separation of the two ends of the D1-brane on the boundary signals non-locality of the spacetime theory (LST) in the UV.}

\item{In the linear dilaton regime (\ie\ $U_0\to \infty$), one can check that $\frac{dL}{dU_0}= \frac{\pi\sqrt{\alpha'}}{2\sqrt{k}U_0^3}+O\left(\frac{1}{U_0^5}\right)>0$. This implies that there is a minimum $L_{\rm{min}}$ satisfying $0<L_{\rm{min}}<\beta_H/2$. This behaviour is puzzling as it implies that $\forall L\in (L_{\rm{min}},\beta_H/2)$, there are two possible configurations of the D1-brane in the bulk (corresponding to two possible values of $U_0$) that minimizes the action. Numerically one can calculate $L_{\rm{min}}=\frac{2.945}{2\pi}\beta_H$. We will see from the discussion that follows, that one of the solution is energetically more favourable than the other. Such double valued behaviour of $L$ also appears in $AdS$ black holes (see \eg\ \cite{Brandhuber:1998bs,Brandhuber:1998er}).}
\end{enumerate}

\begin{figure}[h]
    \centering
    \includegraphics[width=.5\textwidth]{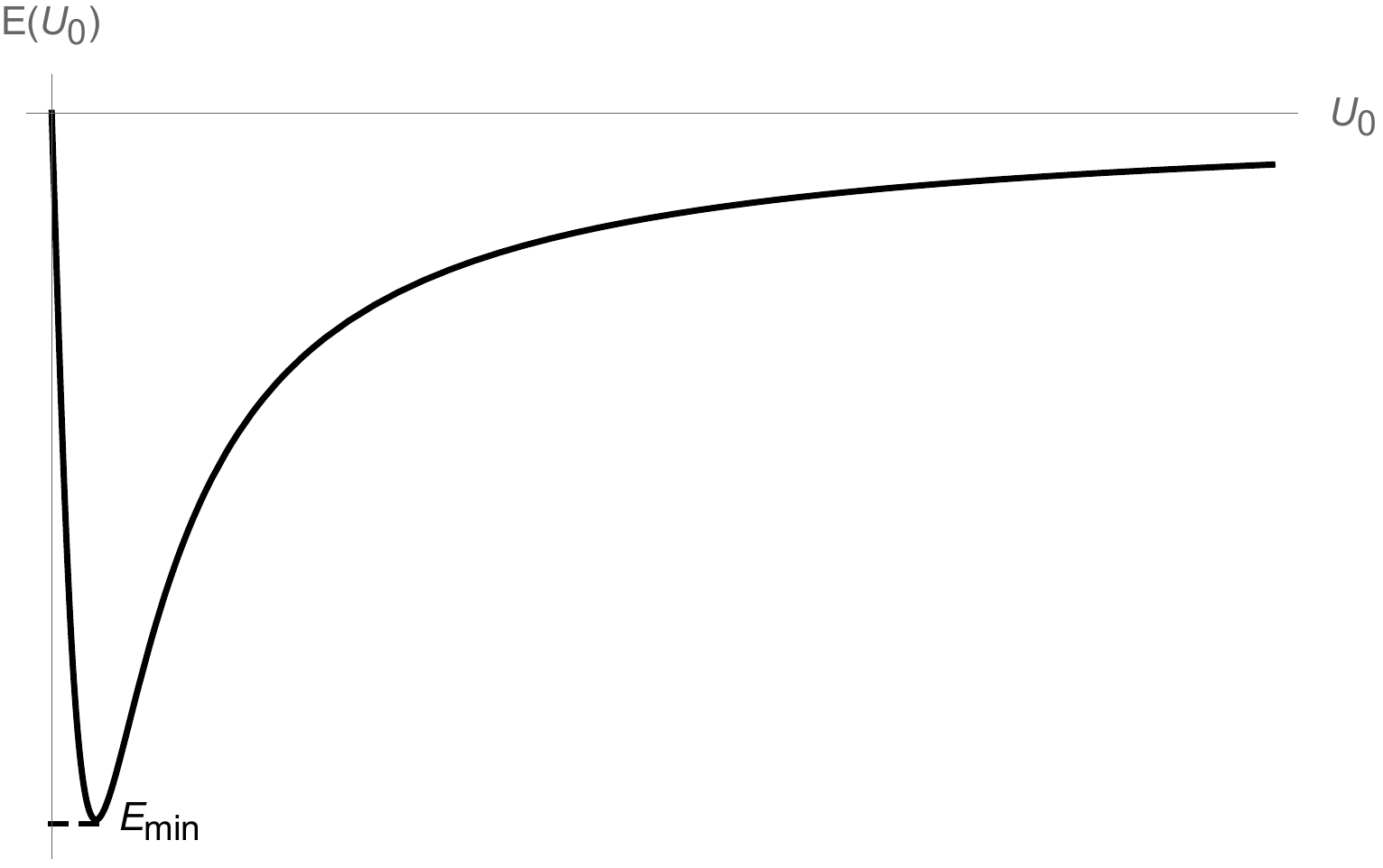}
    \caption{$E(U_0)$ vs $U_0$ in $\mathcal{M}_3$ at zero temperature.}
    \label{M3 t=0 E vs U0}
\end{figure}

\begin{figure}[h]
    \centering
    \includegraphics[width=1.1\textwidth]{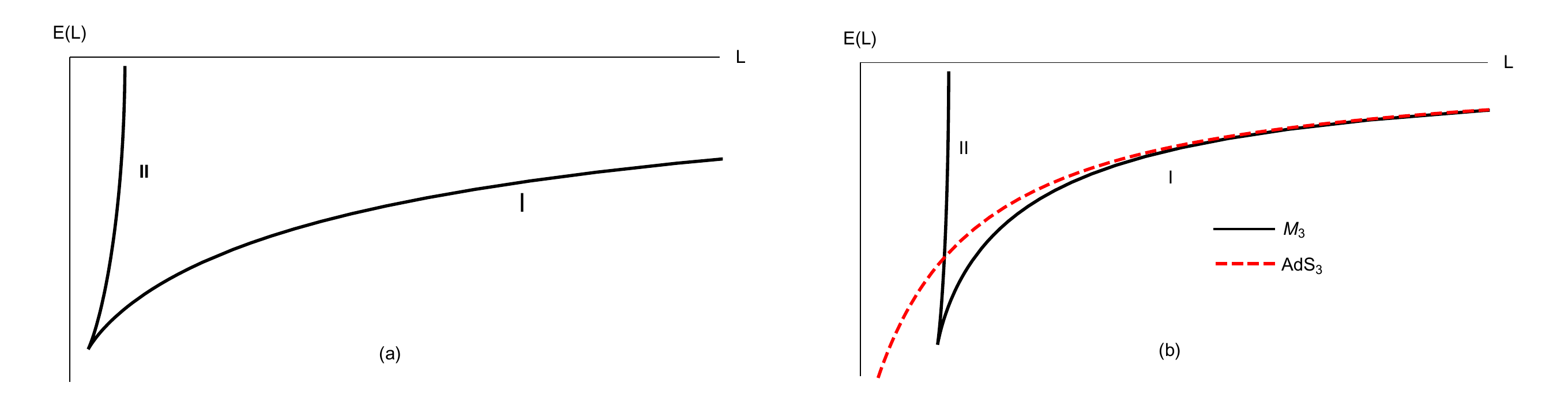}
    \caption{(a) $E(L)$ vs $L$ in $\mathcal{M}_3$ at zero temperature. (b) $E(L)$ vs $L$ $\mathcal{M}_3$ at zero temperature in black and  in $AdS_3$ at zero temperature in dotted red.}
    \label{M3 T=0 E vs L}
\end{figure}

\begin{figure}[h]
    \centering
    \includegraphics[width=1.15\textwidth]{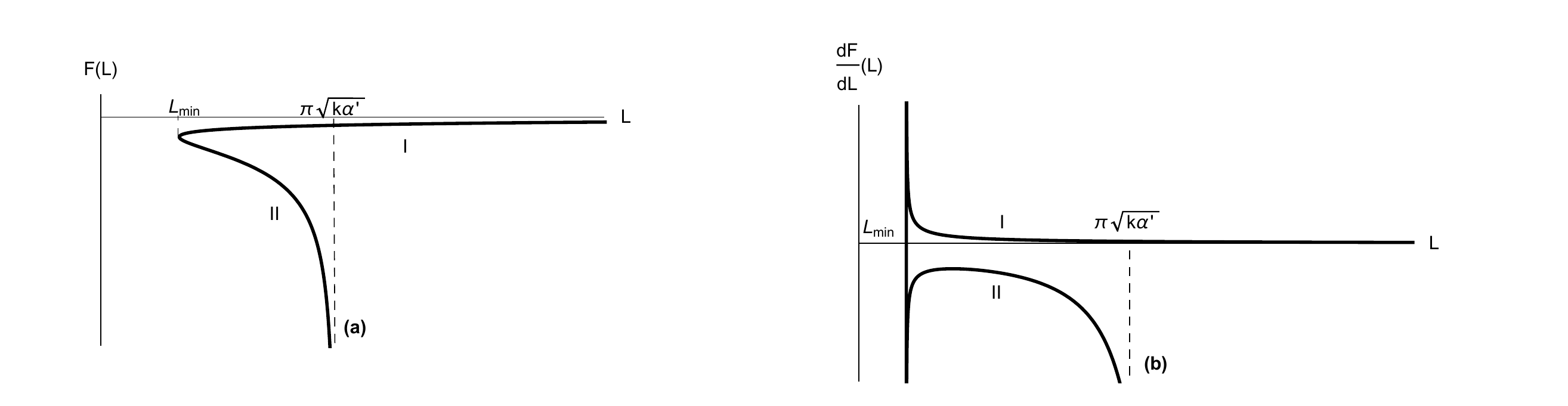}
    \caption{(a) $F(L)$ vs $L$ in $\mathcal{M}_3$ at zero temperature. (b)  $\frac{dF(L)}{dL}$ vs $L$ in $\mathcal{M}_3$ at zero temperature.}
    \label{M3 T=0 F and dF/dL vs L}
\end{figure}

The potential energy of the quark anti-quark system is obtained by plugging \eqref{M3 T=0 constrain} into \eqref{M3 T=0 action}. This integral \eqref{M3 T=0 action}, as expected, would give infinite result because it includes the mass of the D1-brane stretching all the way to infinity. One can regularize the expression by subtracting the disconnected solutions of a pair of D1-branes stretching from $U=0$ to $U=\infty$. So the energy of the system (\ie\ the energy of the connected solution above the pair of disconnected ones), thus obtained, is finite and is given by
\begin{eqnarray}
E&=&\frac{kU_0}{\pi g_s\sqrt{\alpha'}}\left\{\int_1^\infty dy \left( \frac{y^2\sqrt{\frac{1}{U_0^2}+k}}{\sqrt{\frac{1}{U_0^2}+ky^2}\sqrt{\frac{y^4\left(\frac{1}{U_0^2}+k\right)}{\left(\frac{1}{U_0^2}+ky^2\right)}-1}}-1\right)-1\right\}\nonumber \\
&=&\frac{kU_0}{\pi g_s\sqrt{\alpha'}}\Bigg\{-\mathcal{E}\left(\frac{-1}{1+kU_0^2}\right)+\frac{(2+kU_0^2)}{(1+kU_0^2)}\mathcal{K}\left(\frac{-1}{1+kU_0^2}\right)\nonumber \\
&& -\frac{1}{\sqrt{1+kU_0^2}}\left(i\mathcal{K}(-1-kU_0^2)+\mathcal{K}(2+kU_0^2)\right)\Bigg\}.\label{M3 T=0 energy}
\end{eqnarray}
In the deep $AdS_3$ regime (\ie\ $U_0\to 0$), $E$ behaves as
\begin{eqnarray}
E= -\frac{k}{g_s\sqrt{\alpha'}}\left( \mathcal{E}(-1)+\mathcal{K}(2)-\frac{(2-i) \sqrt{\pi } \Gamma \left(\frac{5}{4}\right)}{\Gamma \left(\frac{3}{4}\right)}\right)\frac{U_0}{\pi}+ O(U_0^2),
\end{eqnarray}
whereas in the linear dilaton regime (\ie\ $U_0\to \infty$), $E$ behaves as 
\begin{eqnarray}
E=-\frac{1}{4g_s\sqrt{\alpha'}U_0}+O(U_0^{-2}).
\end{eqnarray}
Figure (\ref{M3 t=0 E vs U0}) shows a schematic variation of $E$ as a function of $U_0$.
The minimum of the energy is given by
\begin{eqnarray}
E_{\rm{min}}&=&-\frac{0.678}{\pi g_s}\sqrt{\frac{k}{\alpha'}}. 
\end{eqnarray}
Both $L(U_0)$ and $E(U_0)$ attain their minima at $U_0=U_0^\ast=\frac{1.144}{\sqrt{k}}$.
 One can obtain $E$ as a function of $L$ by eliminating $U_0$ from \eqref{M3 T=0 length} and \eqref{M3 T=0 energy}. This can be done numerically as shown in figures (\ref{M3 T=0 E vs L}).
Following figure (\ref{M3 t=0 E vs U0}) and (\ref{M3 T=0 E vs L}) few comments are in order:
\begin{enumerate}[{(i)}]
\item{As stated earlier, the energy $E$ along the y-axis in figures (\ref{M3 t=0 E vs U0}) and (\ref{M3 T=0 E vs L})  is the difference between the energies of the connected and disconnected solutions. That $E\leq 0$ for all allowed values of $U_0$ (or equivalently $L$) shows that the disconnected solution is never energetically favourable over the connected one except when $E=0$ where both the solutions are equally favourable. The large and small $L$ configurations are  respectively obtained when: (i) the connected D1-brane is hanging deep in the $AdS_3$ regime (\ie\ $U_0\to 0$) and (ii) when the connected D1-brane is hanging in the linear dilaton regime (\ie\ $U_0\to \infty$). Both these limits agree with the previously known results.}

\item{The double valuedness of $L$ in the $E$ vs $L$ plot in figure (\ref{M3 T=0 E vs L}) together with the kink signals phase transition, in particular second order quantum phase transition which we will analyse, in details, in the discussion that follows. Figure (\ref{M3 T=0 E vs L}) shows that there are two branches namely branch I which we call the $AdS_3$ branch and branch II which we call the linear dilaton (LD) branch. It is obvious from the plot that the branch I is energetically more favourable than branch II.}

\item{As pointed out earlier, the system undergoes second order phase transition at $L=L_C=L_{\rm{min}}$. This phase transition takes place at zero temperature, hence it is a pure quantum phenomenon (see \eg\ \cite{suchdev} for a review on quantum phase transitions). The force $F=-\frac{dE}{dL}$ between the two end points of the D1-brane or equivalently between the quark anti-quark pair is continuous and always attractive (\ie\ $F\leq 0$), but the derivative of the force (\ie\ $\frac{dF}{dL}$) is discontinuous at $L=L_C$ as shown in figure (\ref{M3 T=0 F and dF/dL vs L}). The energy function  has a power series expansion around $L=L_C$ along both branches: $AdS_3$ and linear dilaton, given by
\begin{eqnarray}
AdS_3 \text{ branch: } \ \ E_{I}&=& E_{\rm{min}}+\frac{1}{g_s}\sqrt{\frac{k}{\alpha'}}\sum_{n=1}^\infty \frac{C^I_{\frac{n+1}{2}}}{k^{\frac{n+1}{4}}}\left(\frac{L-L_C}{\sqrt{\alpha'}}\right)^{\frac{n+1}{2}}, \\
\text{LD  branch: } \ \ E_{II}&=& E_{\rm{min}}+\frac{1}{g_s}\sqrt{\frac{k}{\alpha'}}\sum_{n=1}^\infty \frac{C^{II}_{\frac{n+1}{2}}}{k^{\frac{n+1}{4}}}\left(\frac{L-L_C}{\sqrt{\alpha'}}\right)^{\frac{n+1}{2}},
\end{eqnarray}
where $C^{I}$s and $C^{II}$s are coefficients that can be determined numerically.\footnote{Numerical analysis shows that $C^I_{1}=C^{II}_{1}=0.137$ , $C^{I}_{\frac{3}{2}}=-C^{II}_{\frac{3}{2}}=-0.221$, $ C^I_2=C^{II}_2=0.184$, $C^I_{\frac{5}{2}}=-C^{II}_{\frac{5}{2}}=-0.581$, $C^{I}_{3}=C^{I}_{3}=1.985$ and \etc. It could be that the coefficients $C^{I}$s and $C^{II}$s follow a more general pattern: $C^{I}_n=C^{II}_{n}$ and $C^{I}_{n+\frac{1}{2}}=-C^{II}_{n+\frac{1}{2}}$ for all $n\in \{1,2,\cdots \}$.} This shows that the first derivative of the energy function is continuous at $L=L_C$, but the second derivative is discontinuous at $L=L_C$ with both branches diverging as $\pm |L-L_{C}|^{-\frac{1}{2}}$. One can read off the critical exponents of the force $F$ from the first derivative of the energy function $E_I$ and $E_{II}$, which turn out to be $\frac{1}{2}$ along both branches. The point to be noted is that this critical exponent is independent of $k$.
}
\end{enumerate}

\section{Bulk calculation: finite temperature}\label{sec4}

We now generalize the analysis discussed in the previous section to finite temperature. The background \eqref{background} at finite temperature takes the form
\begin{eqnarray}
ds^2&=&-\frac{f_1}{f}dt^2+\frac{1}{f}dx^2+k\alpha'f_1^{-1}\frac{dU^2}{U^2},\nonumber \\
e^{2\Phi}&=&\frac{g_s^2}{kU^2}f^{-1},\label{background T}\\
dB&=&\frac{2i}{U^2}f^{-1}\epsilon_3~ ,\nonumber
\end{eqnarray}
where $f=1+\frac{1}{kU^2}$, $f_1=1-\frac{U_T^2}{U^2}$ and $U_T$ is the radius of the horizon of the black hole in $\MM_3$. The temperature of the black hole is related to the horizon radius $U_T$ via:
\begin{eqnarray}\label{BH temp}
T=\frac{1}{2\pi}\sqrt{\frac{U_T^2}{\alpha'(1+kU_T^2)}}.
\end{eqnarray}
For $\sqrt{k}U_T\ll 1$ the horizon is deep inside the $AdS_3$ regime in $\MM_3$, and to good approximation is describes by BTZ black hole. On the other hand when  $\sqrt{k}U_T\gg 1$ the horizon is deep inside the linear dilaton regime and the solution is well described by the coset $\frac{SL(2,\mathbb{R})\times U(1)}{U(1)}$ \cite{Giveon:2005mi}.

\subsection{Holographic Wilson loop}

\begin{figure}[h]
    \centering
    \includegraphics[width=1.07\textwidth]{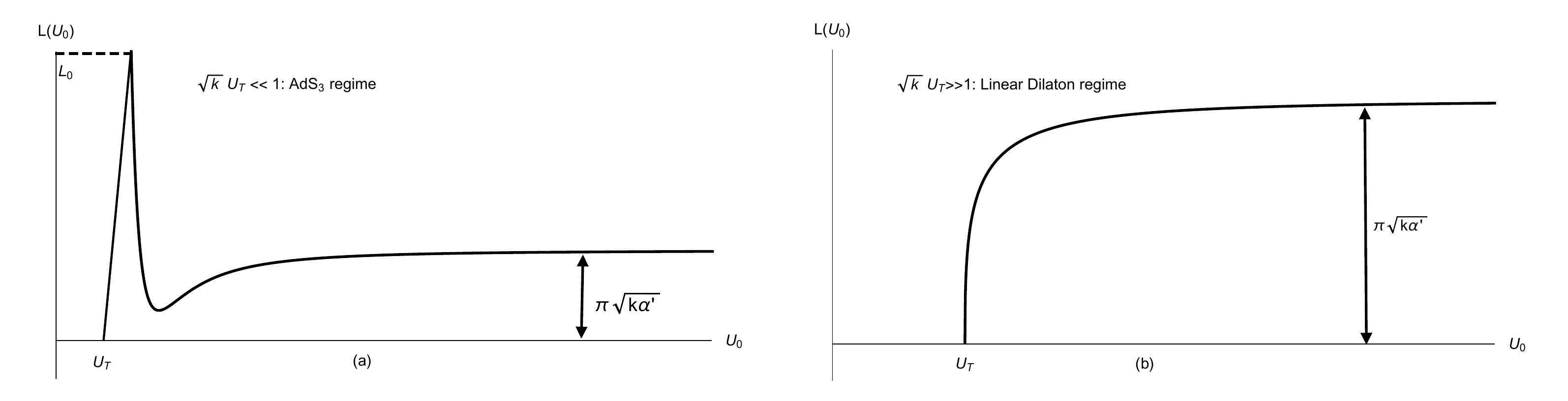}
    \caption{The figure shows a schematic variation of $L$ as a function of $U_0$ in $\mathcal{M}_3$ at finite temperature with: (a) $U_T$ in deep $AdS_3$ regime, (b) $U_T$ in linear dilaton regime. }
    \label{M3 finite T schematic L vs U0}
\end{figure}

The DBI action of the D1-brane hanging in $\MM_3$ at finite temperature is given by
\begin{eqnarray}
S&=&\frac{T_Ek}{2\pi\alpha'g_s}\int dx \sqrt{\frac{U^2(U^2-U_T^2)+\alpha'(kU^2+1)(\partial_x U)^2}{(kU^2+1)}}.
\end{eqnarray}
With the given initial conditions $U(x=0)=U_0$ and $\partial _x U|_{x=0}=0$, the equation of motion is given by
\begin{eqnarray}
\frac{U^2(U^2-U_T^2)}{\sqrt{1+kU^2}\sqrt{U^2(U^2-U_T^2)+\alpha'(1+kU^2)(\partial_x U)^2}}=\sqrt{\frac{U_0^2(U_0^2-U_T^2)}{1+kU_0^2}}.
\end{eqnarray}
Thus, the separation between the two ends of the D1-brane or equivalently the quark anti-quark pair is given by
\begin{eqnarray}
L=2\sqrt{\alpha'}\int_1^\infty dy \frac{\sqrt{\left(\frac{1}{U_0^2}+ky^2\right)}}{\sqrt{y^2(y^2-y_T^2)}\sqrt{\frac{y^2(y^2-y_T^2)\left(\frac{1}{U_0^2}+k\right)}{(1-y_T^2)\left(\frac{1}{U_0^2}+ky^2\right)}-1}},
\end{eqnarray}
where $y=\frac{U}{U_0}$ and $y_T=\frac{U_T}{U_0}$. In the linear dilaton regime (\ie\ $U_0\to \infty$) $L$ behaves as
\begin{eqnarray}
L= \frac{\beta_H}{2}-\frac{\pi}{4U_0^2}(1+kU_T^2)\sqrt{\frac{\alpha'}{k}}+O\left(\frac{1}{U_0^4}\right).
\end{eqnarray}
As expected $L$ approaches $\beta_H/2$ as $U_0\to \infty$  (as in the zero temperature case).
The other extreme limit where $U_0\to U_T$ with $U_T$ held fixed in the $AdS_3$ regime, the separation $L$ goes to 0. Figure (\ref{M3 finite T schematic L vs U0}) shows a schematic variation of $L$ as a function of $U_0$ for the case where the horizon is deep inside $AdS_3$ regime (figure (\ref{M3 finite T schematic L vs U0}(a))) and in linear dilaton regime (figure (\ref{M3 finite T schematic L vs U0}(b))). Unlike the zero temperature case, $L$ here is bounded from above. Let the maximum possible value that $L$ can assume be $L_{\rm{max}}$.  Figure (\ref{M3 finite T schematic L vs U0}(a)) shows that  when the horizon is sitting deep inside the $AdS_3$ regime, the curve has a peak where $\frac{dL}{dU_0}$ vanishes (and $\frac{d^2L}{dU_0^2}<0$). Let this value of $L$ be $L_0$.
  One can infer from figure (\ref{M3 finite T schematic L vs U0}(a)$\&$(b)) that $L_{\rm{max}}=L_0$ when the horizon is deep inside $AdS_3$ regime, and $L_{\rm{max}}=\beta_H/2$ when the horizon is in the linear dilaton regime. $L_0$ is a function of $U_T$ that smoothly goes to $\infty$ as $U_T\to 0$. For $U_T$ in the $AdS_3$ regime and $U_0$ greater than the value where figure (\ref{M3 finite T schematic L vs U0}(a)) develops a peak, the overall behaviour of $L$ as a function of $U_0$ is similar to that of $\mathcal{M}_3$ at zero temperature.

\begin{figure}[h]
    \centering
    \includegraphics[width=1.05\textwidth]{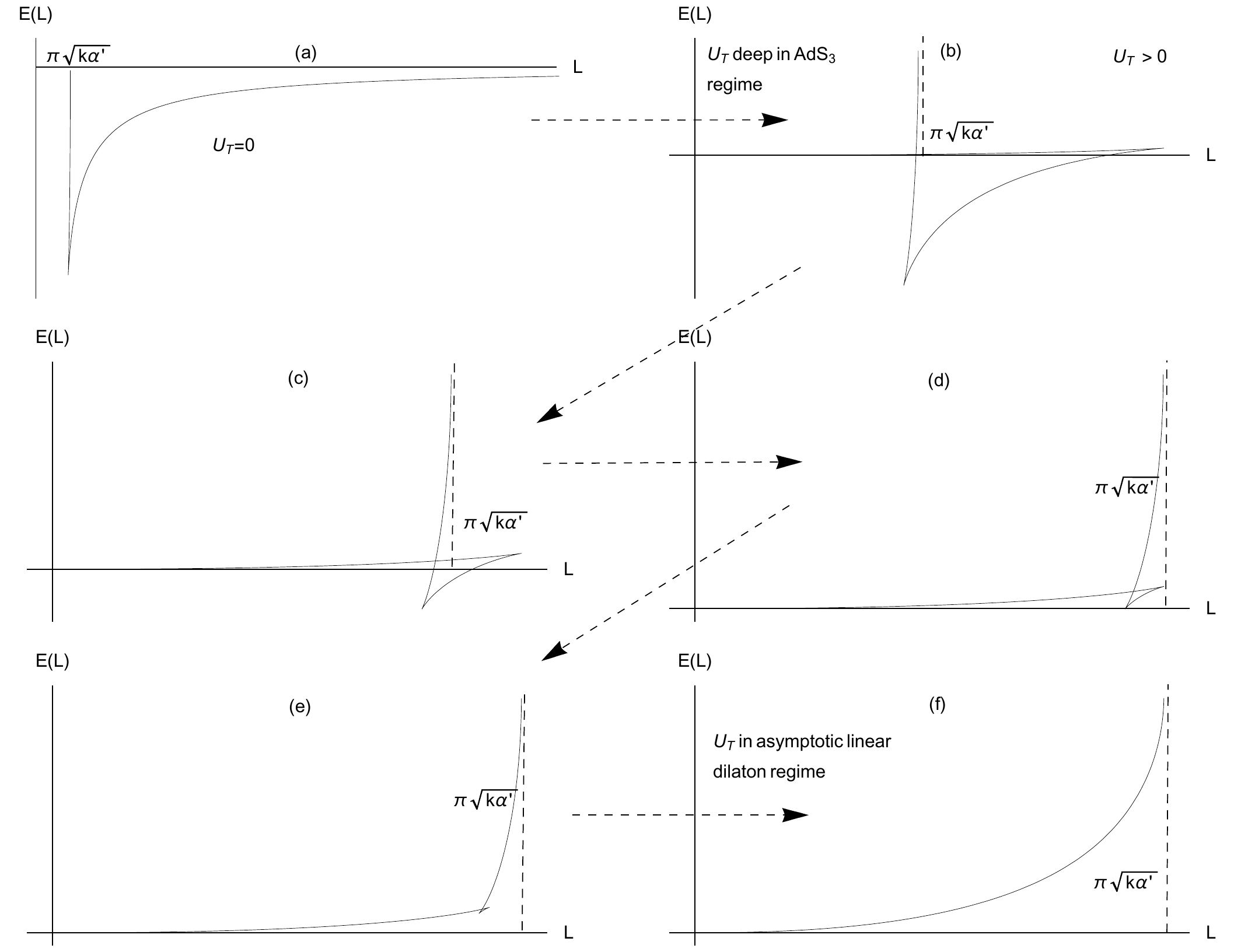}
    \caption{The figure shows the variation of $E$ as a function of $L$ in $\mathcal{M}_3$ at finite temperature. The black dotted arrow shows the variation of $E$ vs $L$ plot as one takes $U_T$ from 0 to the linear dilaton regime. In figure (b) and (c), we see the usual second order quantum phase transition (encountered in the case of zero temperature (figure (a))). Figure (d) shows the transition from connected to disconnected solution, where  thermal fluctuation is comparable to quantum fluctuation. Figure (e) and (f) shows that for $T>T_C$, the disconnected solution is energetically more favourable than the connected ones.}
    \label{M3 finite T E vs L}
\end{figure}

The energy of the quark anti-quark pair above the disconnected solution is given by
\begin{equation}
E=\frac{kU_0}{\pi g_s\sqrt{\alpha'}}\left\{\int_1^\infty dy \left( \frac{\sqrt{y^2(y^2-y_T^2}\sqrt{\frac{1}{U_0^2}+k}}{\sqrt{\frac{1}{U_0^2}+ky^2}\sqrt{\frac{y^2(y^2-y_T^2)\left(\frac{1}{U_0^2}+k\right)}{\left(\frac{1}{U_0^2}+ky^2\right)}-1+y_T^2}}-1\right)-1+y_T\right\}.
\end{equation}
In the limit $U_0\to U_T$, the energy $E$ goes to zero. In the linear dilaton regime (i.e. $U_0\to \infty$), the energy goes to a positive constant (proportional to $U_T$);
\begin{eqnarray}\label{Eld}
E= \frac{k}{\pi g_s\sqrt{\alpha'}}U_T\geq 0.
\end{eqnarray}
Note that the above equation is valid irrespective of the location of the horizon. Equality in \eqref{Eld} holds when the temperature goes to 0. At finite temperature, the asymptotic value of energy (\ie\ $E(U_0\to \infty)$) is positive, implying that the disconnected solution is energetically more favourable than the connected solution in linear dilaton background. Figure (\ref{M3 finite T schematic E vs U0}) shows a schematic variation of $E$ as a function of $U_0$ for the case where the horizon is deep inside $AdS_3$ regime (figure (\ref{M3 finite T schematic E vs U0}(a))) and in linear dilaton regime (figure (\ref{M3 finite T schematic E vs U0}(b))). Figure (\ref{M3 finite T E vs L}) shows the $E$ vs $L$ plot as one takes $U_T$ from zero to the linear dilaton regime. The black dotted arrows  indicate how  $E$ vs $L$ plot changes as $U_T$ increases from 0 to $\sqrt{k}U_T\gg 1$. Based on figure (\ref{M3 finite T schematic E vs U0}) and figure (\ref{M3 finite T E vs L}), the following comments are in order:
 
 \begin{figure}[h]
    \centering
    \includegraphics[width=.9\textwidth]{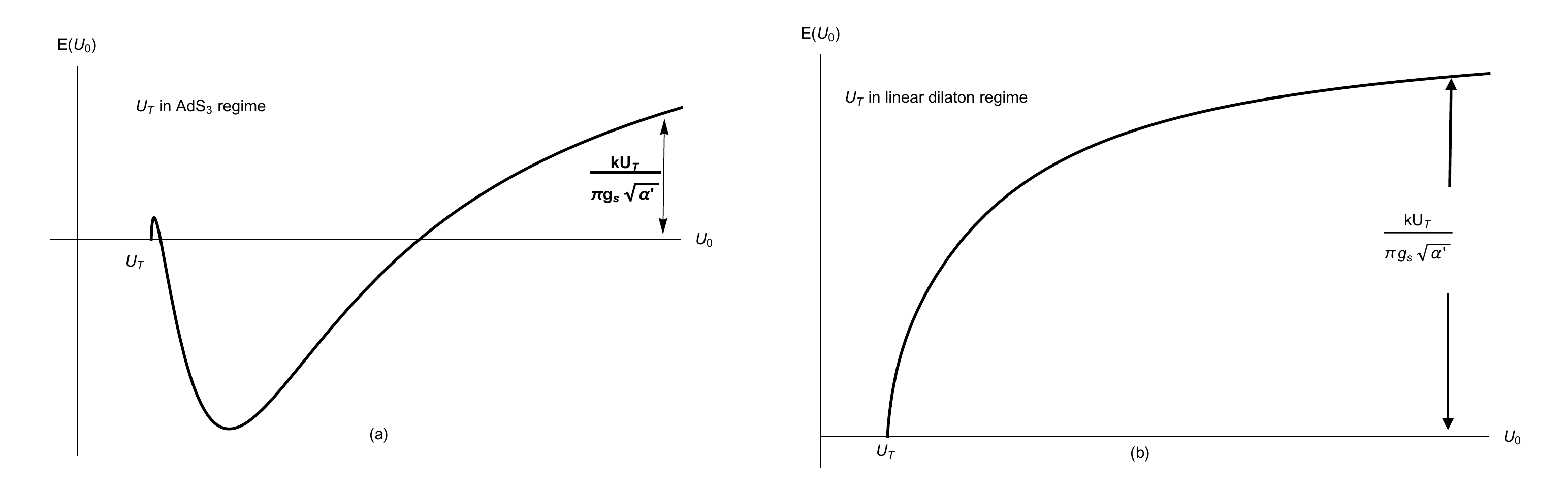}
    \caption{The figure shows a schematic variation of $E$ as a function of $U_0$ in $\mathcal{M}_3$ at finite temperature with: (a) $U_T$ in deep $AdS_3$ regime, (b) $U_T$ in  linear dilaton regime. }
    \label{M3 finite T schematic E vs U0}
\end{figure}

\begin{enumerate}[{(i)}]
\item{In the zero temperature case, we have seen that $L$ is bounded from below by $L_{\rm{min}}$ and unbounded from above. For all $L>L_{\rm{min}}$ there is a connected solution which is energetically favourable over the disconnected solution. As we turn on temperature, we see that this is nolonger the scenario. Finite temperature automatically puts an upper bound on $L$ which goes to infinity as the temperature goes to 0 and to $\beta_H/2$ for very high temperature. Not all allowed values of $L$ produce a negative energy solution. For temperature $T$ less than certain critical value, $T_C$, there is a continuous interval in $L$ where there exist connected solutions (see figure (\ref{M3 finite T E vs L}(b)$\&$(c))). But for $T>T_C$, the size of this continuous interval goes to 0 and there only exist disconnected solutions (see figure (\ref{M3 finite T E vs L}(e)$\&$(f))). Figure (\ref{M3 finite T E vs L}(d)) corresponds to temperature $T=T_C$. For $T<T_C$, there exists a neighbourhood around $T_C$ in which the minimum of the energy (which is the value of the energy at the kink in figure (\ref{M3 finite T E vs L})) behaves as $E_{\rm{min}}\sim |T-T_C|^\alpha$, where the critical exponent $\alpha$ can be computed numerically. This can be realized as a first order thermal phase transition from connected to disconnected solutions. This is the same thermal phase transition encountered in the entanglement entropy analysis in \cite{Chakraborty:2018kpr}. Similar phase transitions also appeared in \cite{Kol:2014nqa}.}

\item{For $T<T_C$, there is the usual second order phase transition (same as the zero temperature case) coming from the discontinuity of $\frac{d^2E}{dL^2}$ at the kink. As in the case of zero temperature, one can calculate the  critical exponents along both branches, $AdS$ and linear dilaton. The sudden disappearance of this second order phase transition for $T>T_C$ can be realized as follows. As we turn on the temperature, thermal fluctuations start competing with quantum fluctuations and as the temperature is increased beyond certain critical value $T_C$, the thermal fluctuations dominate completely over quantum fluctuations and we do not have connected solutions at all. The disconnected solution becomes energetically more favourable.}
\end{enumerate}

\section{Discussion}\label{sec5}
The main goal of this paper is to study the models discussed in \cite{Giveon:2017nie,Giveon:2017myj,Asrat:2017tzd,Chakraborty:2018kpr} through the lens of holographic Wilson loop. We investigated the consequences of non-locality on the quark anti-quark potential energy and the quantum and thermal phase transitions of the theory. The followings are some of the important observations that we have.  
\begin{enumerate}[(i)]
\item{In the UV the separation between the quark anti-quark pair (or equivalently the the two ends of a D1-brane hanging in the bulk) cannot be decreased below a certain minimum distance ($\beta_H/2$) both at zero and finite temperature with connected solution in the bulk. This signals the non-local nature of LST. The non-locality scale observed in the entanglement entropy analysis in $\mathcal{M}_3$ \cite{Chakraborty:2018kpr} is $\beta_H/4$. This difference is probably due to the fact that different probes see different non-locality scales, though each is proportional to $\beta_H$.}

\item{At zero temperature the theory exhibits second order quantum phase transition at $L=L_C$. The force between the quark anti-quark pair exhibits a critical behaviour  with exponent $1/2$ along both branches, $AdS_3$ and linear dilaton. 

This quantum phase transition is not captured by the analysis of entanglement entropy in  $\mathcal{M}_3$ \cite{Chakraborty:2018kpr} implying that this phenomenon is sensitive to how the theory has been is probed. This in particular, is not a robust property of the theory itself, rather it is an artefact of the interaction between the external quark anti-quark pair introduced to probe the theory. }

\item{At finite temperature the theory exhibits first order thermal phase transition at $T=T_C$ from connected to disconnected solution. The same thermal phase transition was observed in \cite{Chakraborty:2018kpr}.  The second order quantum phase transition, encountered in the zero temperature case, persists for $T<T_C$.}
\end{enumerate}
 
It would be interesting to perform perturbative calculation of the expectation of the Wilson loop operator in the $T\bar{T}$  deformed $CFT_2$ and compare with the exact results of section \ref{sec3} and \ref{sec4}. String theory has predictions of different phase transitions in the spacetime theory. It would be nice to understand these transitions from direct field theory analysis.

\section*{Acknowledgements} 

We are grateful to A. Giveon and N. Itzhaki for numerous discussions and throwing in many insightful ideas. We would also like to thank S. Elitzur, D. Kutasov and M. Smolkin for helpful discussions. This work is supported in part by the I-CORE Program of the Planning and Budgeting Committee and the Israel Science Foundation (Center No. 1937/12), and by a center of excellence supported by the Israel Science Foundation (grant number 1989/14).

\end{document}